\documentstyle[12pt]{article}
\begin{document}
\baselineskip 0.9cm
\begin{center}
{\LARGE Dynamical Wormholes and Energy Conditions}
\end{center}


\centerline {\it Anzhong Wang \footnote{e-mail address: wang@on.br}}

\begin{center}
Department of Astrophysics, Observatorio Nacional, Rua General
Jos\'e Cristino $77, 20921-400$ Rio de Janeiro, RJ, Brazil.
\end{center}

\centerline {\it Patricio S. Letelier \footnote{e-mail address:
letelier@ime.unicamp.br} }
\begin{center}
Department of Applied Mathematics-IMECC, Universidade Estadual de Campinas,
$13081-970$ Campinas, SP,  Brazil.
\end{center}

\vspace*{1.cm}

\begin{abstract}
\baselineskip 0.9cm
A class of exact solutions of the Einstein field equations
representing non-static wormholes  that obey the {\em weak and
dominant energy conditions  } is presented. Hence, in principle, these
wormholes can be built with less exotic matter than the static ones.\\


\end{abstract}
\newpage
\baselineskip 0.9cm

The idea of wormholes can be traced back as early as to 1916, when
Flamm [1] studied the just published Schwarzschild solution and found
that its embedding indicates a nontrivial topology. This concept was
further developed by Weyl [2], Einstein and Rosen [3], and Wheeler
[4].  However, it did not receive much attention until recently when
Morris and Thorne [5] pointed out that advanced civilizations might be
able to build a wormhole and to use it in interstellar travels, and
even more to convert it into a time machine and travel back in time.
One of the main obstacles to build such a wormhole is that the matter
needed necessarily violates the  weak energy condition (WEC) [5 - 7].

In this Letter, we present a class of solutions of the Einstein field
equations, which represents non-static wormholes. These solutions are
obtained by the so-called ``cut and paste" approach commonly used in
the recent studies of bubbles [8] as well as wormholes [6, 9].
However, our solutions are differentiated from the already known ones
due to the striking feature that they represent wormholes not
necessarily built with ``exotic" matter. In fact, they satisfy both of
the weak and dominant energy conditions. In this sense they are sharing
the same properties as those studied by the present authors in the
context of the Brans-Dicke theory of gravity [10]. Therefore,  to
construct a wormhole for interstellar and time travels now seems less
forbidden than before.

Before proceeding, we would like to note that the results
presented here do not invalidate the early conclusions of Morris and
Thorne [5] about the static wormholes but rather show the fact that the
properties of dynamic wormholes are quite different from the static
ones. Dynamic wormholes that satisfy the WEC have been also studied recently
in [11, 12]. However, the ones presented here are different from those
given in the above citations and were briefly reported in [13].

The rest of this Letter is organized as follows:  Using the ``cut and
paste" approach, we first construct a class of solutions of the
Einstein field equations, which represents a spherically symmetric {\em
non-static } bubble.  Then, we  analyze these solutions and show that
the bubble acts as the throat of a wormhole which connects two
asymptotically flat universes without horizons and singularities.
Finally, we show that such constructed bubbles  satisfy the
weak and dominant (but not strong) energy conditions [14].

To begin with, let us consider the Schwarzschild solution of the Einstein
vacuum equations
\begin{equation}
ds^2 = f d T^2 - f^{-1} dR^2 - R^2 d^2 \Omega,
\end{equation}
where the function $f$ is defined as usual $f \equiv 1 - 2m/R$, and $d^2
\Omega \equiv d^2 \theta + \sin^2 \theta d^2 \phi$. The coordinates have the
ranges $-\infty < T < + \infty, 0 \leq R < + \infty, 0 \leq \theta \leq
\pi$, and $0 \leq \phi \leq 2 \pi$. Following Ref. [10], we perform
the following coordinate transformations
$$ T = M(t-\psi) + N(t+\psi), R = U(t-\psi) + V(t+\psi), \; {\rm for}
\; \psi \geq 0 \eqno{\rm (2a)}$$
$$ T = M(t+\psi) + N(t-\psi), \quad R = U(t+\psi) + V(t-\psi), \; {\rm for}
\; \psi \leq 0 \eqno{\rm (2b)}$$
where $M, N, U$ and $V$ are functions of the class $C^4$ in
their indicated arguments in the sense defined in [14],
and $t$ and $\psi$ are new coordinates
with $-\infty < t, \psi < + \infty$.
In terms of $t$ and $\psi$, the metric (1) becomes
\setcounter{equation}{2}
\begin{equation}
ds^2 = Fdt^2 + 2G dtd\psi - H d^2 \psi - R^2 d^2 \Omega ,
\end{equation}
where $F \equiv f T^2_{,t} - f^{-1} R^2_{,t}, \;\; G \equiv f T_{,t}
T_{,\psi} - f^{-1} R_{,t}  R^2_{,\psi}$, and $H \equiv f^{-1}
R^2_{,\psi} - f T^2_{,\psi}$. From Eqs. (2) one can see that in each of
the two regions, $\psi \geq 0$ and $\psi \leq 0$, the metric (3) is
locally isometric to the one given by Eq. (1). In other words, they are
related to each other by the well-defined coordinate transformations
(2a) and (2b). However, across the hypersurface $\psi = 0$, the
functions $T$ and $R$ given by Eqs. (2) are only continuous functions
of $\psi$ (class $C^0$). Consequently, the metric coefficients $F, G$,
and $H$ are discontinuous functions of $\psi$ across $\psi=0$.  In order
that the Einstein equations hold in the sense of distributions [15] the
metric coefficients need to be at least $C^0$  across the
hypersurface.  It is easy to show that the condition
\begin{equation}
\dot{M}^2 - \dot{N}^2 = \frac{(U+V)^2}{(U+V-2m)^2} (\dot{U}^2-\dot{V}^2),
\qquad (\psi = 0),
\end{equation}
where an overdot denotes the ordinary differentiation with respect
to $t$,  guarantees  the continuity of  the metric coefficients across
$\psi = 0$.

For  our purpose, we find  sufficient to consider only the special
solutions of Eq. (4) with $\dot{M}^2 - \dot{N}^2 = \mu,
\;\; M = AN + B$, and $U = a V + b$, where $\mu, A, B, a$ and $b$ are
arbitrary constants. When $\mu = 0$,  the solutions reduce to the ones
considered by Visser [9], which represent static wormholes and therefore
belong
to the class studied by Morris and Thorne [5].  The weak energy
condition is violated in those solutions.  In the rest of the Letter we
shall focus our attention on the solutions with $\mu \neq0$. The
integration of Eq. (4) yields
\begin{eqnarray}
&& N(t) = \varepsilon_1 \mu_1 t + N_0 , \nonumber \\
&& \varepsilon_2 \mu_2 t = V(t) + \frac{2m}{1+a} \ln (\hat{R}(t) - 2m)
+ V_0,
\end{eqnarray}
where $N_0$ and $V_0$ are integration constants, $\mu_1 \equiv
\sqrt{\mu / (A^2-1)},\;\; \mu_2 \equiv \sqrt{\mu / (a^2-1)}, \;\;
\varepsilon_1, \varepsilon_2 = \pm 1$, and $\hat{R}(t) \equiv R(t,
\psi=0) = (1+a)V(t) + b.$ Once Eq. (4) is solved, the metric
coefficients of  (3) are in turn fixed in terms of $t$ and $\psi$. As
mentioned previously, such solutions are isometric to the Schwarzschild
metric (1) in the regions $\psi \geq 0$ and $\psi \leq 0$,
respectively. Therefore, in these two regions the energy-stress tensor
vanishes. For metrics of the class $C^0$ the energy-stress tensor
is distribution valued [15] across the hypersurface $\psi =0$.  In the
present case we find
\begin{equation}
T_{\mu \nu} = \tau_{\mu \nu}\delta (\psi) = \{\sigma u_\mu u_\nu - \tau
(\theta_\mu \theta_\nu + \phi_\mu
\phi_\nu) \} \delta (\psi) ,
\end{equation}
where $\delta(\psi)$ denotes the usual Dirac delta function, $u_\mu =
\sqrt{F} \delta^t_\mu, \;\; \theta_\mu = \hat{R}(t) \delta^\theta_\mu$,
and $\phi_\mu = \hat{R}(t)\sin \theta \delta^\phi_\mu$. The function
$\sigma$ denotes the surface energy density of the bubble and $\tau$
the tensions in the tangential directions  measured by an observer with
the four-velocity $u_\mu$, and are given respectively by
\begin{equation}
\sigma = \frac{\sigma_0}{\hat{R}(t)} , \quad \tau =
\frac{\sigma_0 (\hat{R}(t)-m)}{2\hat{R}(t) (\hat{R}(t) - 2m)},
\end{equation}
where  $\sigma_0 = 2\varepsilon_2 (1+A) /[\mu_2 (a-A)]$ is a constant
[16].  We conclude that our solutions represent a spherical bubble
connecting two regions, each of which is locally isometric to part of
the Schwarzschild space-time. Since the radius of the bubble
$\hat{R}(t)$ is time-dependent, the bubble in the present case is not
static. From Eq. (5) we see that $\hat{R}(t) \geq 2m$. That is, the
bubble is  always greater than  or equal to the Schwarzschild sphere,
where the equality holds only at $t = + \infty$.  In order
that  the hypersurface $\psi=0$ be non-spacelike, we need to impose the
condition $\mu (a-A)(1+a)(1+A) > 0$.

The dynamics of the bubble can be studied using the kinematical
quantities
\begin{equation}
\hat{R}' = \beta
\left(\frac{\hat{R}(t)-2m}{\hat{R}(t)}\right)^{\frac{1}{2}}, \quad
\hat{R}'' = m \beta^2 \frac{1}{\hat{R^2}(t)} ,
\end{equation}
where $\beta \equiv \varepsilon_{2}\mu_2 (1+a)
\sqrt{(1-a)(1-A)/[2\mu(a-A)]}$, and a prime denotes differentiation
with respect to the proper time measured by the observer who is at rest
relative to the bubble. The above equations show that in the present
case the bubble either expands $(\beta > 0)$ or collapses $(\beta  <
0)$, depending on the choice of the free parameters. Now we shall
consider the following representative cases:

(a) $\varepsilon_2 = -1,\;\; 0 < a < 1, \;\; 0 < A < 1,\;\;
A > a,\;\; b = 2m$,
and $V_0 = 0$. Then, we find
\begin{eqnarray}
Exp (-\mu_2 t) &=& [(1+a)V(t)]^{\frac{2m}{1+a}} Exp[V(t)] , \nonumber\\
R(t, |\psi|) &=& a V (t-|\psi|) + V(t+|\psi|) + 2m, \nonumber\\
\hat{R}(t) &=& (1+a) V(t) + 2m .
\end{eqnarray}
{}From the above equation, it is easy to show that at any moment, say,
$t=t_1$, we always have $R(t, |\psi|) \rightarrow + \infty$ as $\psi
\rightarrow \pm \infty$, and that $R(t, |\psi|) \geq 2m$ for any $t$
and $\psi$, where the equality holds only when $t = +\infty$.  Thus, in
the present case the bubble acts as the {\em throat of a wormhole }
that connects two asymptotically flat Schwarzschild universes. Note
that in [9] non-static wormholes were also studied.  However, our
solutions are different from those.  This can be seen clearly by the
following considerations. From Eqs.(4), (5) and (9) we find that
$[\partial R(t, |\psi|)/\partial |\psi|] |_{\psi = 0} = -(1-a)\mu_{2} f
< 0.$ Therefore, in the present case the radius of the wall is
initially decreasing as away from it. Then, according to the results
obtained in Ref. 8, the surface energy density of the bubble is
positive [cf. Eq.(7)]. When the radial coordinate reaches its minimum,
say, $R_{min.}$, which is always greater than or equal to $2m$ (Again,
the equality holds only when $t = + \infty$), it starts to increase. As
$|\psi| \rightarrow + \infty$, we have $R(t, |\psi|) \rightarrow +
\infty.$ In [9], the case with $[\partial R(t, |\psi|)/\partial |\psi|]
|_{\psi = 0} > 0$ was studied. As a result, the surface energy density
is always negative and violates all the energy conditions [8, 9]. It
should be noted that usually [17] the signs of the angular component of
the extrinsic curvature tensor of the wall  were taken as the criterion
to classify the spatial topology of the  spacetimes. This is correct
for the static case.  However, when the spacetime is time-dependent, it
is true only in the neighborhood of the bubble. The global topology of
the spacetime could be completely different.  This fact has been
noticed in [18] quite recently and the present solutions provide
another example.

On the other hand, from Eq.(9) we find that $\hat{R}(t) \rightarrow
+\infty$ as $t \rightarrow - \infty$, and $\hat{R}(t)  \rightarrow 2m$
as $t \rightarrow + \infty$.  That is, the corresponding solutions
represent collapsing wormholes. The wormhole throat starts to collapse
at $\hat{R}(t) = \infty$ and ends at $\hat{R}(t)=2m$. From Eq. (8) one
can show that to complete this process, the throat takes an infinite
proper time. Consequently, a space adventurer will have enough time to
pass through the throat of the wormhole from one asymptotically flat
region to the other before the radius of the throat shrinks to $2m$,
where  the event horizon is developed. A distinguish feature of this
wormhole is that  it consists of matter that satisfies the weak and
dominant energy conditions [14] as long as $\hat{R}(t)  \geq 3m$.  The
latter can be seen clearly from Eq. (7). Note that, although  these two
energy conditions are satisfied, the strong one is not, and the throat
of the wormhole  is still gravitational repulsive, since the
``Newtonian mass" of the throat (that is proportional to $(\sigma -
2\tau)$) is less than or equal to zero. The equality holds only at the
initial point $t = - \infty$. The repulsive character is needed in
order to keep the throat open [5].

As mentioned before, the above results do not contradict to the ones
presented in [5]. To show this, let us consider the embedding of our
solutions in the  three dimensional Euclidean space $ds^2 = dZ^2 + dR^2
+ R^2 d^2 \phi$. Following Ref. [5], we find $(dZ/d\psi)^2 = g(t,
\psi)$, where $g(t, \psi)$ changes signs from point to point.  That is,
in the present case our solutions can not be embedded in a
3-dimensional Euclidean space and pictured as an ordinary Euclidean
curved surface.  From Ref. [5] we  see that the violation of the weak
energy condition is tightly related to such a three dimensional
embedding.  Recall that not any two dimensional metric  can be embedded
into a three dimensional Euclidean space. Classical examples are the
Moebius strip and the Gauss-B\'olyai-Lobachevski metric $ds^2 =
(1+r^2)^{-1} dr^2 + r^2 d\phi^2$ [19].

(b) $\varepsilon_2 = 1, \;\; a > 1,\;\; A > 1,\;\; a > A, \;\; b = 2m,
$ and $V_0 = 0$. We find that $Exp (\mu_2 t) =
[(1+a)V(t)]^{\frac{2m}{1+a}}
 Exp[V(t)]$, and that $R$ and $\hat{R}(t)$ are still given by Eq. (9).
As in the previous case, one can show that these solutions represent
expanding bubbles, which connects two asymptotically flat universes.
The bubble expands from $\hat{R}(t) = 2m$ to $\hat{R}(t) = + \infty$ by
taking infinite proper time. Again, when $\hat{R}(t) \geq 3m$, the
bubble satisfies the weak and dominant energy conditions.

{}From Eq. (7) we can see that when $\hat{R}(t)$ is approaching $2m$, the
tensions in the tangent directions of the bubble tend to infinite.
Thus, in the course of the collapse of the bubble, as described in the
first case, it is not difficult to imagine that the bubble will explode
due to the enormous tensions before its radius shrinks to $2m$. By
properly arranging the parameters, the explosion could happen as early
as wanted. After the explosion, the material may recompose and form
another wormhole, the later evolution of which follows more or less the
same process as described by the solutions of Case (b).

In this Letter we have shown that traversable wormholes can be built
out of matter that satisfies the weak and dominant energy conditions. The
violation of the strong energy condition nowadays does not seem to be
a very serious drawback (Recall that cosmic bubbles and domain walls
formed in the early Universe do not satisfy this condition either).

The recent studies of wormholes usually fall into two different
directions.  One is concerned with the energy conditions [7], and the
other is concerned with the vacuum polarization due to the quantum
effects [20, 21]. To the first, one can see that even it can be shown
that the WEC is preserved at the quantum level for the generic cases,
the existence of wormholes can not be ruled out. As shown above, they
can exist even in the classical level without violating the WEC.  To
the second, Hawking [20] and Visser [21] argued that, when the vacuum
polarization effects are taken into account,  one might finally show
that such a building of a traversable wormhole is impossible, although
Thorne and others [22] seem to defend the opposite opinion.  The
considerations of the latter are out of scope of this Letter and will
be discussed somewhere else.

\vspace{1.cm}

\noindent{\bf Acknowledgments}

Part of the work was done when one of the authors (A.W.) was visiting
the Department of Applied Mathematics, UNICAMP. He thanks the
hospitality of the Department. He also gratefully acknowledges the
financial assistence from FAPESP and CNPq.

\end{document}